\documentclass[%
reprint,
superscriptaddress,
 nofootinbib,
 amsmath,amssymb,
 aps,
 physrev,
floatfix
]{revtex4-2}

\usepackage{graphicx}% Include figure files
\usepackage{dcolumn}% Align table columns on decimal point
\usepackage{hyperref}

\let\tempcorr\corresponds % to avoid error in mathabx
\let\corresponds\relax
\usepackage{mathabx}
\let\corresponds\tempcorr

\usepackage[dvipsnames]{xcolor}
\usepackage[x11names]{xcolor}
\usepackage[normalem]{ulem}

\begin{document}

\preprint{}

\renewcommand{\thempfootnote}{\arabic{mpfootnote}}
\newcommand{\Nadd}{N_\text{add}}
\newcommand{\Naddup}{N_\text{add}^\text{up}}
\newcommand{\Nadddown}{N_\text{add}^\text{down}}
\newcommand{\Nout}{N_\text{out}}
\newcommand{\Noutup}{N_\text{out}^\text{up}}
\newcommand{\Noutdown}{N_\text{out}^\text{down}}
\newcommand{\Naddm}{N_\text{add,m}}
\newcommand{\Naddtilde}{N_\text{add,e\kern-0.1em /\kern-0.15em o}}
\newcommand{\Naddintegrated}{\bar{N}^\text{up}_\text{add}}
\newcommand{\Ncorr}{N_\text{add,corr}}
\newcommand{\GammaT}{\Gamma_\text{T}}
\newcommand{\Gammae}{\Gamma_\text{e}}
\newcommand{\Gammao}{\Gamma_\text{o}}
\newcommand{\Gammain}{\Gamma_\text{in}}
\newcommand{\gammam}{\gamma_\text{m}}
\newcommand{\etat}{\eta_\text{t}}
\newcommand{\etaapp}{\eta}
\newcommand{\etaM}{\eta_\text{M}}
\newcommand{\kappae}{\kappa_\text{e}}
\newcommand{\kappaeext}{\kappa_\text{e,ext}}
\newcommand{\kappao}{\kappa_\text{o}}
\newcommand{\kappaoext}{\kappa_\text{o,ext}}
\newcommand{\kappain}{\kappa_\text{in}}
\newcommand{\kappainext}{\kappa_\text{in,ext}}
\newcommand{\omegam}{\omega_\text{m}}
\newcommand{\omegas}{\omega_\text{s}}
\newcommand{\fm}{f_\text{m}}
\newcommand{\fs}{f_\text{s}}
\newcommand{\hstiffe}{N_\text{det,e}(\fs)}
\newcommand{\hstiffo}{N_\text{det,o}(\fs)}
\newcommand{\nm}{\bar n_\text{m}}
\newcommand{\nth}{n_\text{th}}
\newcommand{\no}{n_\text{om}}
\renewcommand{\ne}{n_\text{em}}
\renewcommand{\ae}{a_\text{e}}
\newcommand{\be}{b_\text{e}}
\newcommand{\neffo}{\bar n_\text{o}}
\newcommand{\neffe}{\bar n_\text{e}}
\newcommand{\nmine}{n_\text{min,e}}
\newcommand{\nmino}{n_\text{min,o}}
\newcommand{\Ae}{\mathcal{A}_\text{e}}
\newcommand{\Ao}{\mathcal{A}_\text{o}}
\newcommand{\Ain}{\mathcal{A}_\text{in}}
\newcommand{\Thr}{\Theta }
\newcommand{\Capac}{\mathcal{C}_\text{ub}}
\newcommand{\SiN}{Si$_3$N$_4$}
\newcommand{\ebps}{ebps}

\title{High-throughput bidirectional electro-optic transduction\\assessed with a practical quantum capacity}

\author{M. D. Urmey}
\affiliation{JILA, National Institute of Standards and Technology and the University of Colorado, Boulder, Colorado, 80309, USA}
\affiliation{Department of Physics, University of Colorado, Boulder, Colorado, 80309, USA}

\author{S. Dickson}
\affiliation{JILA, National Institute of Standards and Technology and the University of Colorado, Boulder, Colorado, 80309, USA}
\affiliation{Department of Physics, University of Colorado, Boulder, Colorado, 80309, USA}

\author{K. Adachi}
\affiliation{JILA, National Institute of Standards and Technology and the University of Colorado, Boulder, Colorado, 80309, USA}
\affiliation{Department of Physics, University of Colorado, Boulder, Colorado, 80309, USA}

\author{S. Mittal}
\affiliation{JILA, National Institute of Standards and Technology and the University of Colorado, Boulder, Colorado, 80309, USA}
\affiliation{Department of Physics, University of Colorado, Boulder, Colorado, 80309, USA}

\author{L. G. Talamo}
\affiliation{JILA, National Institute of Standards and Technology and the University of Colorado, Boulder, Colorado, 80309, USA}
\affiliation{Department of Physics, University of Colorado, Boulder, Colorado, 80309, USA}

\author{A.~Kyle}
\affiliation{JILA, National Institute of Standards and Technology and the University of Colorado, Boulder, Colorado, 80309, USA}
\affiliation{Department of Physics, University of Colorado, Boulder, Colorado, 80309, USA}

\author{N. E. Frattini}
\affiliation{JILA, National Institute of Standards and Technology and the University of Colorado, Boulder, Colorado, 80309, USA}
\affiliation{Department of Physics, University of Colorado, Boulder, Colorado, 80309, USA}

\author{S.-X. Lin}
\affiliation{JILA, National Institute of Standards and Technology and the University of Colorado, Boulder, Colorado, 80309, USA}
\affiliation{Department of Physics, University of Colorado, Boulder, Colorado, 80309, USA}

\author{K. W. Lehnert}
\affiliation{Department of Physics, Yale University, New Haven, Connecticut 06511, USA}

\author{C. A. Regal}
\affiliation{JILA, National Institute of Standards and Technology and the University of Colorado, Boulder, Colorado, 80309, USA}
\affiliation{Department of Physics, University of Colorado, Boulder, Colorado, 80309, USA}

\date{\today}

\begin{abstract}
A microwave-optical transducer of sufficiently low noise and high signal transfer rate would allow entanglement to be distributed between superconducting quantum processors reliably within the lifetimes of their quantum memories.
We report the continuous operation of an efficient transducer at high enough bandwidth to realize kHz signal transfer rates, in downconversion as well as upconversion, with noise levels approaching that required for quantum information transfer.
To define the significance of our advance and clarify the multifaceted performance required for quantum networking tasks, we derive a broadband quantum channel capacity that bounds the maximum rate at which quantum information can be sent through realistic finite‑bandwidth thermal‑loss channels. 
This capacity serves as a comprehensive measure of transducer capability and provides insight into the relative importance of disparate performance metrics. 
We find that the broadband capacity depends on the throughput---defined as the product of efficiency, bandwidth, and duty cycle---and on the added noise.
We present measurements of a membrane-based opto-electromechanical transducer with high throughput of 7~kHz and at an input-referred added noise of 3 photons in both upconversion and downconversion, demonstrating that bidirectional transducer capacities comparable to superconducting qubit decay rates are within reach.  
In downconversion, throughput of this magnitude at the few-photon noise level is unprecedented, marking an improvement of nearly four orders of magnitude over previous work. 
\end{abstract}

\maketitle

\section{Introduction}
A low-noise transducer bridging microwave and optical frequencies would allow the deterministic generation of arbitrary optical quantum states for applications in metrology, the storage and manipulation of optical quantum information, and the networking of remote superconducting quantum processors. 
Reducing input-referred added noise $\Nadd$ to a level far below one photon is imperative for such quantum state manipulation. 
At the same time, it is useful to increase the rate of transduction to reach signal rates that are compatible with the lifetimes of logical qubits that process the transduced quantum information. 
Current state-of-the-art lifetimes in superconducting architectures are approximately 1~ms using physical qubits~\cite{ganjam2024surpassing, bland2025millisecond} and error-corrected bosonic encodings~\cite{sivak2023real}, and 300~$\mu$s using logical encodings of many physical qubits~\cite{google2023suppressing}. 
Finite qubit lifetimes place constraints not only on the transducer bandwidth $B$ (which must be fast enough to transduce quantum information within the time qubits can faithfully store it), but also on the overall signal throughput, in order to support an entangled bit rate much faster than decoherence rates. 
Signal throughput is further reduced from $B$ by less-than-unity transducer efficiency $\eta$ and duty cycle $D$.
Accordingly, we identify a transducer's throughput $\Thr$~\cite{brubaker2022optomechanical,zhao2025quantum} quantitatively with the efficiency-bandwidth-duty-cycle product $\Thr = \eta B D$, as a way to characterize the total rate of transduction.
Roughly, it gives the expected rate that single photon signals are detected at the output of a transducer, when sent at a repetition rate of $B$.
An experimental result demonstrating low-noise transduction at sufficient $\Thr$ to reliably entangle superconducting qubits within their lifetimes remains outstanding.

In practice, there is often a tradeoff between $\Nadd$ and $\Thr$, as increasing $B$, $\eta$, or $D$ typically involves increasing the average drive power, which can lead to heating-induced noise.
Because they each can have different impacts on the potential to distribute entanglement, prior experimental work has treated $\Nadd$, $B$, $\eta$, and $D$ as separate performance metrics. 
As a result, the quantum transduction research community has lacked a framework to combine all the relevant parameters into a practical metric for assessing transducer performance. 

An additional consideration is how a transducer performs in each direction of transduction, which dictates the specific applications and protocols it can enable.
For instance, long-range quantum networks dominated by loss will likely benefit from using heralding protocols, in which entanglement infidelity due to inefficiency is mitigated at the cost of transmission rates and greater noise sensitivity~\cite{zeuthen2020figures}.
Such a protocol favors a network topology with transducers operating in upconversion, from microwave to optical frequencies, to take advantage of mature optical single-photon-detection technologies~\cite{marsili2013detecting}. 
To date, most all demonstrations (and strictly all quantum demonstrations) have taken place in upconversion.
A fully developed transducer platform, alternatively, would be capable of operating bidirectionally, achieving comparable performance in both directions of frequency conversion. 
Transducers capable of downconversion (from optical to microwave frequencies) are needed to unlock the full set of proposed protocols for optically distributing entanglement between superconducting nodes. 
For instance, transducers operating in downconversion would be useful for generating microwave entanglement from squeezed optical light~\cite{kyle2023optically}. 
Further, deterministic state transfer protocols~\cite{cirac1997quantum, axline2018demand} are possible if efficiencies are high, and require one transducer operating in upconversion and one in downconversion.
Finally, inherently bidirectional performance allows the efficient use of additional ports (e.g.\ monitoring reflection) to enable entanglement-preserving protocols with an effective $\Nadd<1$, even in transducers that demonstrate inherent entanglement-breaking $\Nadd>1$ when operated as a direct one-mode channel~\cite{higginbotham2018harnessing, zhang2018quantum, rau2022entanglement}.
Published work to date has operated more favorably in upconversion than in downconversion, and there are only a few examples that quantify downconversion performance~\cite{sahu2022quantum, jiang2023optically}.

\begin{figure}
    \centering
    \includegraphics{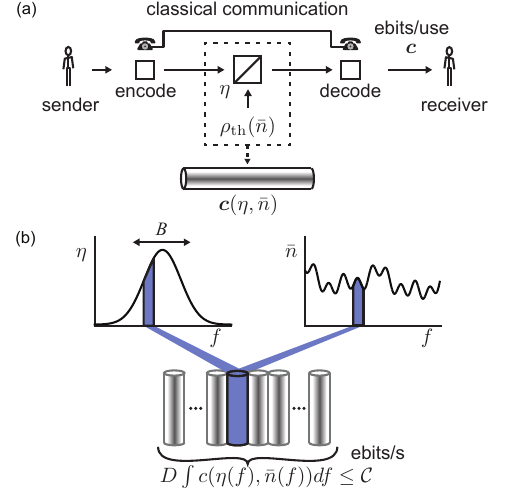}
    \caption{Quantum channel with finite bandwidth and thermal noise. 
    (a) The two-way classically assisted quantum capacity $\boldsymbol{c}$ of a channel with efficiency $\eta$ and thermal noise $\rho_\text{th}(\bar n)$ depends on $\eta$ and the average thermal occupation $\bar n$.
    (b) A channel with noise spectrum $\bar n(f)$ and limited bandwidth $B$, described by a frequency-dependent efficiency $\eta(f)$, can support a broadband quantum capacity $\mathcal{C}$, in units of ebits/s (\ebps). 
    Duty cycle $D$ lower than unity also impacts $\mathcal{C}$.
    Therefore, in addition to the noise of the channel, $\eta$, $B$, and $D$, the constituents of throughput $\Thr$, affect $\mathcal{C}$.
    Because $\boldsymbol{c}$ is superadditive, integrating $\boldsymbol{c}(f)$ places a lower bound on $\mathcal{C}$.}
    \label{fig:cartoon}
\end{figure}

In this work, we present measurements of a transducer with high throughput $\Thr$  exceeding 7~kHz, and noise performance close to input-referred added noise $\Nadd = 1$, expressed in units of photons, or photons per second per Hz of bandwidth, in upconversion ($\Naddup = 2.6$) as well as downconversion ($\Nadddown = 3.6$). 
While similar throughputs have been achieved in a device with quantum-enabled performance in upconversion~\cite{zhao2025quantum}, in downconversion, our device increases throughput by nearly four orders of magnitude over previous work at the threshold of quantum-enabled performance~\cite{sahu2022quantum}.
Our platform achieves such performance due to its robustness to strong optical pumping~\cite{brubaker2022optomechanical}, as well as significant improvement to the electrical circuit that we show here.

In order to provide a single, practical metric to assess the utility of a transducer in quantum networks from its noise and throughput, in this work, we analytically integrate upper bounds to the two-way quantum channel capacity. 
We make the experimentally motivated assumption of transducers with Lorentzian transmission and spectrally flat, thermal input-referred added noise~\cite{brubaker2022optomechanical, zhao2025quantum, higginbotham2018harnessing, sahu2022quantum, jiang2023optically, mirhosseini2020superconducting, meesala2024non,  kumar2023quantum, xie2025scalable, weaver2024integrated}. 
The resulting broadband quantum channel capacity describes an upper bound on the rate at which quantum information can be transmitted through realistic transducers.
Furthermore, this capacity lends theoretical justification to the adoption of the efficiency-bandwidth-duty-cycle product as a measure of a transducer's throughput~\cite{brubaker2022optomechanical, zhao2025quantum}, and illustrates the significance of our experimental result.

\section{Deriving a practical capacity for quantum transduction}

For practical applications of a quantum transducer, performance improvements to both noise and throughput are needed in the field.
Both metrics are important for the overall quantum information transfer rate;
a noisier yet higher-throughput transducer could achieve a comparable total entanglement generation rate at a given fidelity as a lower-throughput and lower-noise one. 
For example, a higher-throughput transducer could generate a larger quantity of entangled resource states within quantum memories' lifetimes, which could then be consumed in an entanglement distillation protocol in order to match the fidelity permitted by a lower-noise transducer~\cite{duan2000entanglement,mele2025maximum}. 
We would like a single metric that quantifies a transducer's ability to transmit quantum information---one which accounts for the effects of both throughput and additive noise.

A channel's quantum capacity is the theoretical limit to the rate at which quantum information may be transmitted through it in the presence of unavoidable errors (such as additive noise and loss), and is therefore a good metric to use in evaluating a transducer's performance for quantum processing tasks.
In the context of general-capability quantum nodes linked by noisy and lossy channels, realistic communication protocols will likely use local operations and two-way classical communication (LOCC) to process transmitted quantum information.
For example, heralding protocols are necessary for recovering nonzero rates when losses exceed 50\% and one-way deterministic state transfer becomes impossible~\cite{caruso2006degradability,wolf2007quantum}.
When LOCC operations are allowed, the upper bound to the rate is referred to as the two-way classically assisted quantum capacity $\boldsymbol{c}$, which quantifies the number of qubits that can be faithfully communicated, or ebits of entanglement that can be distributed, on average per attempted use of the channel (Fig.~\ref{fig:cartoon}.(a))~\cite{bennett1996mixed}.
The two-way capacity $\boldsymbol{c}$ is therefore the relevant capacity for many applications in sensing and networking, where LOCC is a free resource and the errors are dominated by loss and thermal noise.
Such bosonic thermal-loss channels can be modeled as a beamsplitter of transmissivity $\eta$ with a thermal state of $\bar{n}$ mean photons incident on the loss port (Fig.~\ref{fig:cartoon}.(a)).

A realistic channel that models transduction has finite bandwidth $B$, so the relevant capacity is per-time rather than per-use~\cite{wang2022quantum,gandotra2026quantum}.
To quantify this capacity, such a channel can be described using a continuous collection of thermal-loss channels given by $\eta(f)$ and $\bar{n}(f)$, provided the thermal baths are spectrally uncorrelated (Fig.~\ref{fig:cartoon}(b)).
However, the two-way quantum capacity $\boldsymbol{c}(\eta, \bar{n})$ of thermal-loss channels is currently not known, nor is it known to be additive~\cite{noh2020enhanced}.
Thus integration over frequency $f$ only gives a lower bound on the broadband capacity $\mathcal{C}\ge\int\boldsymbol{c}(\eta(f), \bar{n}(f))df$, in units of ebits/s or \ebps, with equality holding only for additive-capacity channels (e.g.\ pure loss channels which are always either degradable or anti-degradable)\cite{caruso2006degradability}. 
While calculating $\boldsymbol{c}$ remains an open problem, fortunately, known capacity upper bounds $\boldsymbol{c}_\text{ub}$ for thermal-loss channels are additive~\cite{pirandola2017fundamental,wilde2017converse}.
Integrating these additive upper bounds then yields a broadband capacity upper bound $\Capac = \int\boldsymbol{c}_\text{ub}df\ge \mathcal{C}$ applicable to finite-bandwidth spectrally-uncorrelated thermal-loss channels.
Note that any protocol of the form of Fig. 1 that provides an explicit encoder and decoder has an asymptotic rate that gives a lower bound on the capacity.

In deriving an experimentally relevant expression for $\Capac$ for transducers, we assume transmission $\eta(f)$ to be Lorentzian, which is typically the case in experimental demonstrations. 
For the thermal bath spectrum, we assume frequency-independent added noise $\Nadd=\bar{n}(f)\big(1-\eta(f)\big)/\eta(f)$.
These choices for the frequency dependence of $\eta$ and $\Nadd$ describe transducers whose thermally occupied modes are upstream of a bandwidth-limiting coupling, for instance, a transducer with a thermally occupied microwave mode operating in upconversion. This description applies to all experimental upconversion work shown and cited in this work.
(In practice, the lower energy of microwave photons leads to departure from the assumed noise spectrum in downconversion, from non-negligible heating of e.g.\ the microwave waveguide in Refs.~\cite{hease2020bidirectional} and \cite{sahu2022quantum}, or the microwave resonator in this work and Refs.~\cite{brubaker2022optomechanical} and \cite{jiang2023optically}, see Fig.~\ref{fig:downconversion}(c) below.)
Additionally, for transducers with limited duty cycle, we make the simplifying assume that operating time is sufficiently long that $\boldsymbol{c}(f)$ is not adversely affected, i.e.\ that the transducer's $\mathcal{C}$ is merely penalized by a factor of $D$ for the time it is inoperational.

Under these realistic assumptions, we find that $\Capac$ is approximately linear in the throughput $\Thr$ for small efficiencies, and is given by
\begin{equation}\label{eq:integratedCapacitySmallT}
    \Capac(\Nadd, \Thr) \approx 
    \frac{\pi\; \Thr}{2\ln(2)}
    \bigl( 1 -\Nadd
    +\Nadd \ln(\Nadd)
    \bigr)  
\end{equation}
for $\Nadd\leq1$ and zero otherwise (see Appendix~\ref{app:integrating-capacity} for details).
In the small-$\etaapp$ limit, $\Capac$ depends only on $\Thr$ and $\Nadd$, consolidating the effects of these two separate performance metrics.
Though this approximation is pessimistic for channels with high efficiency, it remains accurate to within 17\% for $\etaapp\leq0.5$ (encompassing all transducer demonstrations currently), and to within 3\% for $\etaapp\leq0.1$.
For clarity, we here point out that this result may conflict with intuition derived from deterministic state transfer protocols, which are effective only in the limit of high efficiency and thus their performance isn't captured by $\Thr$ alone. 

An alternative to using a device as a direct transduction primitive (Fig.~\ref{fig:cartoon}) is to intrinsically generate a microwave-optical squeezed resource for distributing entanglement (i.e.\ by using a blue-detuned pump)~\cite{ barzanjeh2012reversible, zhong2024microwave}, followed by teleportation to transmit the desired quantum information. 
Transducers operated in this microwave-optical squeezing mode have been used to infer continuous variable entanglement between the microwave and optical output modes~\cite{sahu2023entangling}, and to prepare microwave-optical entangled photon pairs demonstrated by optically heralded single microwave photons~\cite{meesala2024quantum}.
A bidirectional quantum transducer is in principle capable of generating symmetric microwave-optical entanglement at rates comparable to its transduction throughput.
More theoretical investigations are needed in order to connect our broadband quantum capacity to an analogous broadband relative entanglement entropy of microwave-optical entanglement generation in order to compare the performance of devices operating as direct transducers or entanglement generators.

\section{Experimental platform}

\begin{figure}
    \centering
    \includegraphics{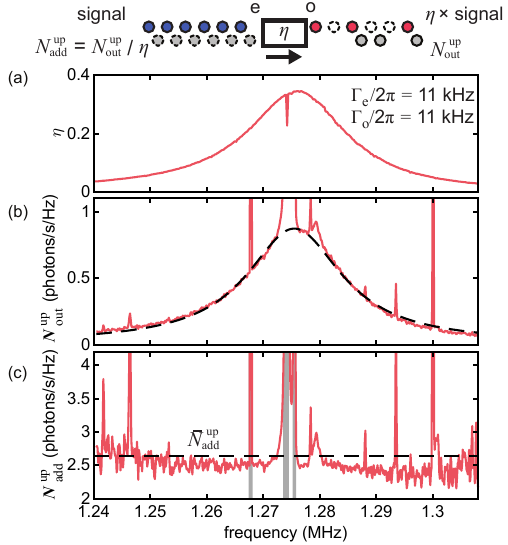}
    \caption{Upconversion transducer performance with $B=22$~kHz. 
    The efficiency $\etaapp$ quantifies the fraction of the incident microwave signal (blue circles) that is transduced to optical frequencies (red circles), rather than lost (empty dashed circles). 
    The upconversion input-referred added noise $\Naddup$ (gray dashed circles) compares the noise rate measured at the output $\Noutup$ (gray circles) to a potential signal at the input. 
    (a)~Transduction efficiency $\etaapp$ vs.\ signal detuning from pump $f$. 
    (b)~Output-referred upconversion noise spectrum. 
    (c)~Input-referred added noise spectrum in upconversion.
    The grayed-out regions indicate frequency bands contaminated by noise from the high thermal occupation of optomechanically stiff modes. 
    Excluding these regions and integrating the noise across the bandwidth of the transducer yields $\Naddintegrated=2.6$ (horizontal dashed line).
    The traces shown are the average of four individual measurement sets taken under the same operating conditions.
    }
    \label{fig:spectrum}
\end{figure}
\begin{figure}[t!]
    \centering
    \includegraphics{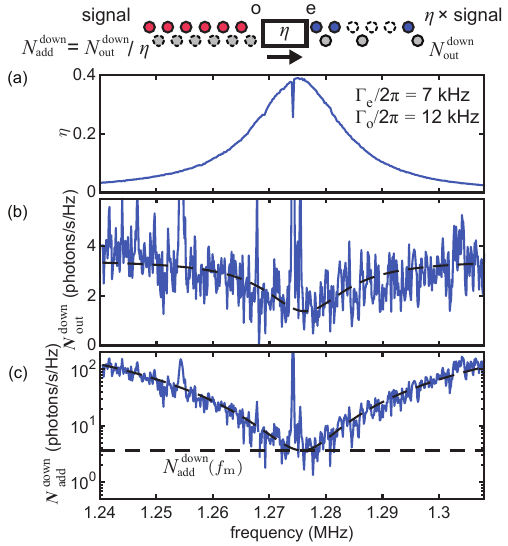}
    \caption{
    Downconversion transducer performance with $B=19$~kHz. 
    (a)~Transduction efficiency $\etaapp$ vs.\ signal detuning from pump $f$. 
    (b)~Downconversion output-referred noise spectrum.
    The broadband noise contribution extending beyond the damped linewidth of the transduction mode is due to unwanted additional occupation of the LC circuit resulting from the microwave pump.
    Destructive interference between this microwave noise and the mechanical motion it drives results in noise squashing (Eq.~\ref{eq:Nadddown}). 
    (c)~Input-referred added noise spectrum in downconversion. 
    Broadband circuit noise divided by the frequency-dependent $\eta$ results in $\Nadddown$ increasing with detuning from the center of the transduction bandwidth.
    Additionally, noise squashing near resonance suppresses the noise level near $\fm=1.275$~MHz.
    }
    \label{fig:downconversion}
\end{figure}

We use a doubly-parametric membrane-based transducer, of similar design to Refs.~\cite{brubaker2022optomechanical, urmey2024stable}, which has demonstrated the highest achieved microwave-to-optical transduction efficiency, and continuous operation.
We have enabled an increase in the operating bandwidth by two orders of magnitude relative to our most recent previous transduction demonstration~\cite{brubaker2022optomechanical} with the following improvements: the silicon nitride \SiN{} in the device is annealed to reduce pump-dependent circuit loss~\cite{mittal2024annealing} and noise (Appendix~\ref{app: technical noise}), and the membrane is ``pinned" in place with a silicon post to more reproducibly set the transducer's capacitor spacing to just 200 nm~\cite{urmey2024quantum,mittal2024enhanced}. 
We have also reduced the noise of the optical and microwave pumps with additional filtering (Appendix~\ref{app: technical noise}).
We emphasize here that the measurements presented in this work serve as a first demonstration that annealing \SiN{} ameliorates microwave-pump induced noise in microwave circuits whose electric field penetrates the \SiN{} (Appendix~\ref{app: technical noise}).

In our transducer, the mechanical mode of the \SiN{} membrane with resonant frequency $\fm =1.275$~MHz couples an optical Fabry-P\'erot cavity and a microwave-frequency superconducting LC circuit via the optomechanical and electromechanical interactions~\cite{andrews2014bidirectional}. 
Electrical and optical pumps red-detuned by $\fm$ from their respective resonators enhance these interactions, coupling the mechanical mode to the microwave (optical) field at angular loss rate $\Gammae$ ($\Gammao$), proportional to the pump power.
Imperfect suppression of the Stokes sideband of the transduction mode leads to gain $\mathcal{A}=\Ae\Ao$, where $\Ae$ ($\Ao$) is the portion of the gain due to the electromechanical (optomechanical) interaction, which is undesirable because it is accompanied by additional noise~\cite{andrews2014bidirectional,caves1982quantum}.
Our device operates with sufficient sideband resolution to limit gain to $\mathcal{A}=1.1$.

The efficiency $\etaapp=\mathcal{A}\etat$ includes this gain factor and gives the fraction of the incident signal that exits the transducer, where $\etat=\etaM(4\Gammae\Gammao/\GammaT^2)$ is the ideal transduction efficiency of a device operating with perfect optomechanical and electromechanical sideband resolution~\cite{brubaker2022optomechanical}.
The maximum achievable efficiency $\etaM$ is determined by loss in the electromagnetic resonators and an optical mode matching factor.
In this device we achieve efficiencies of $\etaapp=0.4$, calibrated by the procedure used in Ref.~\cite{higginbotham2018harnessing}.
The transduction bandwidth $B = \GammaT/2\pi$ is determined by the total loss rate of the mechanical mode~\cite{andrews2014bidirectional}, including the pump-enhanced loss rates, $\GammaT=\Gammae + \Gammao +\gammam$, where $\gammam\ll\Gammae,\Gammao$ is the intrinsic loss rate of the transduction mode, as can be seen in the measurements of $\eta(f)$ in Figs.~\ref{fig:spectrum} and~\ref{fig:downconversion}. 

\section{Upconversion transducer performance}

The measured bidirectional efficiency $\etaapp(f) = \sqrt{\epsilon_\text{PL}}|S_\text{eo}(f)||S_\text{oe}(f)|/|S_\text{ee}(f)||S_\text{oo}(f)|$ is shown in Fig.~\ref{fig:spectrum}(a), 
where e.g.\ $S_\text{oe}$ denotes the scattering parameter governing a signal incident on the electrical port ``e" and detected from the optical port ``o", and $\epsilon_\text{PL}$ is a factor quantifying the spatial mode matching of the optical pump and local oscillator beam used for heterodyne readout~\cite{andrews2014bidirectional}. 
The efficiency follows a Lorentzian lineshape corresponding to the susceptibility of the damped mechanical transduction mode. 

The upconversion output-referred noise spectrum $\Noutup=N_\text{det,o}/\xi_\text{o}$ (Fig.~\ref{fig:spectrum}(b)) corresponds to the noise measured at the optical heterodyne detection $N_\text{det,o}$ in units of photons/s/Hz, divided by the total detection efficiency $\xi_\text{o}$ between the output of the transducer and the detected voltage (see Ref.~\cite{brubaker2022optomechanical} for additional details). 
Here we measure $\xi_\text{o}$ by calibrating the optomechanically measured spectrum in phonon units using sideband asymmetry thermometry.
In addition to the thermomechanical noise following the Lorentzian profile of the damped, optomechanically compliant mode that we use for transduction, the spectrum shows structure due to multiple optomechanically stiff modes, characterized by significant motion in the massive substrate that makes them resistant to optomechanical damping, and whose large thermally driven motion is written onto the optically measured output spectrum. 
The prominence of stiff modes in this generation of device is due to an error in the fabrication geometry, which can be addressed in subsequent mechanical designs (Appendix~\ref{app: stiff modes}).
The Lorentzian fit (dashed curve) excludes the stiff modes, to provide a guide to the eye.
The  fit departs from the measurement in the tails because of a small asymmetry in the spectrum due to interference with another, highly damped membrane mode with a 950~kHz resonant frequency.

\begin{figure*}[t!]
    \centering  \includegraphics{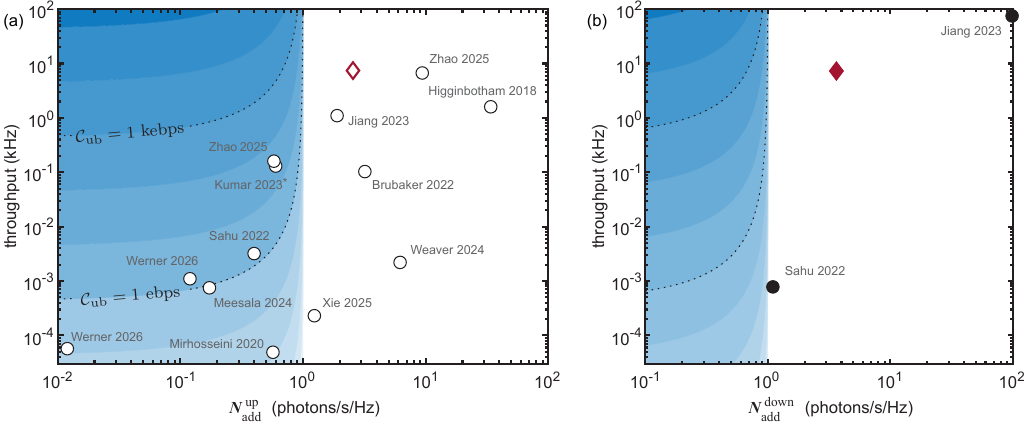}
    \caption{The broadband quantum channel capacity of transducers in upconversion (a) and downconversion (b).
    Measurements presented in this work (maroon diamonds) are compared with other reported results (circles, labeled by first author, see Appendix~\ref{app: prior work})~\cite{mirhosseini2020superconducting, meesala2024non,sahu2022quantum, kumar2023quantum, xie2025scalable, weaver2024integrated, brubaker2022optomechanical, jiang2023optically, higginbotham2018harnessing, zhao2025quantum, werner2026electro}.
    We include all published work that reported sufficient information to characterize direct transduction with $\Nadd<100$.\protect\footnotemark[4]
    Contours indicate $\Capac$ as a function of a transducer's throughput and $N_\text{add}$, assuming $\etaapp\ll 1$ and frequency-independent $\Nadd(f)$. 
    Successive darker shades of blue mark factor-of-ten increases in $\Capac$.
    Surpassing the $\Capac = 1$~kebps line is required to support the transduction of one qubit per 1~ms superconducting register lifetime, on average.
    }
    \label{fig:comparison}
\end{figure*}

Whereas previous work has focused on the noise level at the center of the transduction bandwidth, in order to more transparently convey the potential to resolve a transduced signal, we compute the frequency-dependent input-referred added noise $\Naddup$ (Fig.~\ref{fig:spectrum}(c)) by referring the optical output spectrum shown in Fig.~\ref{fig:spectrum}(b) to the input of the device by dividing by an interpolation of the frequency-dependent transduction efficiency in Fig.~\ref{fig:spectrum}(a):
$\Naddup(f) = \Noutup(f)/\etaapp(f)$.
Neglecting the features due to the stiff modes, the $\Naddup$ spectrum is relatively flat over the frequency band relevant for the transducer. 
Notching out the frequencies corresponding to the grey regions in Fig.~\ref{fig:spectrum}(c) and integrating the total contributed noise, we obtain an average added noise of $\bar{N}^\text{up}_\text{add}=2.64(26)$ over the bandwidth of the transducer (Appendix~app: stiff modes), close to the value of the input referred added noise at the transducer's maximum efficiency, as determined by the fit to the $\Noutup$ noise peak, $\Naddup(\fm)=2.56(26)$. 
The uncertainty in both $\Naddintegrated$ and $\Naddup(\fm)$ is dominated by the statistical uncertainty of the fit parameters in the sideband asymmetry thermometry measurement used to extract $\xi_\text{o}$, and systematic uncertainty in the value of $\epsilon_{PL}$, both of which are used to calibrate this $\Naddup(f)$ spectrum.
If we were to physically implement the sharp filter to remove the impact of the stiff modes in a quantum network, the ringing of the narrow filter would only modestly impact the throughput and noise performance (Appendix~\ref{app: stiff modes}). 
Depending on the application, shaping the signal pulses to avoid problematic frequency regions or using a selective demodulation envelope in detection are also viable solutions to such frequency-dependent noise profiles.
Of course, the more practical solution to the additional noise of the stiff modes is to eliminate them altogether.

The operating parameters used in Fig.~\ref{fig:spectrum} achieve $B=22$~kHz, which represent an improvement in bandwidth by approximately two orders of magnitude compared with the device measured in Ref.~\cite{brubaker2022optomechanical}.
Though it may be expected that such an increase in optomechanical and electromechanical damping rates would come at the cost of a commensurate increase in the noise, we measure similar baseline noise levels, i.e.\ neglecting the noise caused by the stiff modes. 
This improvement in the robustness of our transducer to strong driving is due to improvement of the single-photon electromechanical coupling rate $g_\text{e}$,
reduction in phase noise of both optical and microwave pumps from additional filtering (Appendix~\ref{app: technical noise}), 
and annealing the \SiN{} to remove two-level-system defects~\cite{mittal2024annealing} that were a source of internal loss and, as we find here (Appendix~\ref{app: technical noise}), noise in the microwave circuit. 

\section{Downconversion transducer performance}

Fig.~\ref{fig:downconversion} shows downconversion measurements analogous to those in Fig.~\ref{fig:spectrum}. 
Whereas the damping rates used in Ref.~\cite{brubaker2022optomechanical} would have required prohibitively long averaging times to resolve an electromechanical spectrum with the relative frequency resolution shown here, the orders-of-magnitude greater transduction bandwidths used in this work make such electromechanical measurements tractable in a reasonable amount of time.

Unlike in upconversion, wideband noise extends beyond the bandwidth of the transduction mode due to occupation of the microwave resonator (Fig.~\ref{fig:downconversion}(b)).
This additional noise is associated only with driving the circuit strongly with the microwave pump; we do not observe additional occupation associated with the strong optical pump.
Because the microwave noise field also drives the mechanical mode, there is interferometric cancellation at the mechanical frequency leading to squashing measured at the output port~\cite{jayich2012cryogenic, safavi2013laser}. 
The deficit of noise seen at this output is instead routed out the optical port optomechanically.
In Fig.~\ref{fig:downconversion}(c), in addition to $\Nadddown$ being suppressed from squashing near mechanical resonance, $\Nadddown$ increases with further detuning from the transduction mode because the relatively broadband noise from microwave circuit occupation is divided by the frequency-dependent conversion efficiency. 
Unlike upconversion performance, the $\Nadddown(f)$ spectrum is frequency dependent, and the impact of the off-resonant noise would depend on the details of the network in which the transducer is embedded and the protocol used. 
We therefore for simplicity report downconversion performance using the noise at peak efficiency $\Nadddown(\fm)$ (horizontal dashed line), determined from the minimum of the fit to $\Noutdown(f)$ referred to the optical input (dashed curve), in line with previous work~\cite{higginbotham2018harnessing,hease2020bidirectional,brubaker2022optomechanical}, and analogous to applying additional filtering on $\Noutdown$~\cite{sahu2022quantum}.  
We find $\Nadddown(\fm)=3.6(9)$, comparable to $\Naddup(\fm)$.
That our transducer exhibits bidirectional performance is remarkable, while the difference between the spectra in upconversion and downconversion alludes to an underlying directional asymmetry we explore in Sec.~\ref{sec:asymmetry}.

\footnotetext[4]{Though Refs.~\cite{meesala2024non} and \cite{jiang2023optically} were operated in an microwave-optical squeezing configuration, the characterization in those works contained sufficient detail that direct transduction performance could be ascribed (Appendix~\ref{app: prior work}). Note for these works $\Nadddown>100$. 
We additionally note that Ref.~\cite{kumar2023quantum} (point with asterisk) demonstrated transduction between optical and millimeter wave frequencies.}

The increase in statistical noise in the spectrum measurements shown in Fig.~\ref{fig:downconversion} relative to those in Fig.~\ref{fig:spectrum} is due to the difference between the optical measurement efficiency $\xi_\text{o}=0.4$ and the measurement efficiency of the HEMT amplifier and microwave detection chain $\xi_\text{e}=0.01$.
The reduced signal-to-noise ratio makes calibrating $\xi_\text{e}$ electromechanically via sideband asymmetry impractical, so instead we take advantage of our simultaneous optomechanical and electromechanical measurement of the prominent stiff modes (Appendix~\ref{app: mw readout cal}).

\section{Assessing transducer performance}

Having quantified the bidirectional noise and throughput of our transducer, we can now use $\Capac$ to assess its performance.
Following Ref.~\cite{zhao2025quantum}, we summarize the parameter space of transducer performance by plotting throughput $\Thr$ vs.\ $\Nadd$ in Fig.~\ref{fig:comparison}, a representation supported by the dependence of $\Capac$ on these two quantities.
Contours of $\Capac$ give an intuition for the relative importance of throughput and added noise in the quantum-enabled regime of $\Nadd <1$: although reducing $\Nadd$ gives very sharp improvement while $\Nadd\approx1$, further reductions in noise yield diminishing returns. 
Meanwhile, because $\Capac$ is linear in $B$ and $D$, and at least linear in $\eta$, improving a channel's throughput is always beneficial.

In Figs.~\ref{fig:comparison}(a) and (b), respectively, we put our device's upconversion and downconversion performance measurements in the context of other results in the field by plotting $\Naddup(\fm)$ and $\Nadddown(\fm)$ vs $\Thr$.
In upconversion, we see that the throughput we achieve at the noise levels we measure are consistent with the best achieved in the field~\cite{zhao2025quantum}, though further decreases in pump power do not lead to quantum operation in our current device due to its elevated mechanical loss rate (see Sec.~\ref{sec:asymmetry} and Fig.~\ref{fig:performance}(a)).

In downconversion (Fig.~\ref{fig:comparison}(b)), our device demonstrates a leap in achieved performance, with an increase in throughput of a factor of 9400 compared with previous work~\cite{sahu2022quantum}.
This orders-of-magnitude improvement is accompanied by only a factor-of-three greater $\Nadd$.

\section{Directional asymmetry in transducers}
\label{sec:asymmetry}

Fig.~\ref{fig:comparison} highlights a notable performance difference in the field between upconversion demonstrations and downconversion demonstrations.
This discrepancy may initially be surprising given that a transducer operated continuously at maximum efficiency would ideally be bidirectional: it would achieve equal performance in upconversion and downconversion for lossless cavities.
However, details of the noise baths involved can motivate operating the transducer with diminished efficiency to reduce added noise in one direction at the expense of higher noise in the other.

To explore asymmetry in noise performance between upconversion and downconversion, we distinguish expressions for $\Naddup(\fm)$ and $\Nadddown(\fm)$, making the simplifying assumptions of lossless cavities operating in the resolved sideband regime $\kappae,\kappao\ll4\fm\cdot2\pi$ with perfect optical mode matching:
\begin{align}
    \Naddup(\fm) &= 
    \underbrace{ \frac{\nth\gammam}{\Gammae}+ \ne +  \frac{\no\Gammao}{\Gammae} }_{\Naddm}
    + \underbrace{ \frac{\neffo\GammaT^2}{\Gammae\Gammao} }_{\Naddtilde}
    - \underbrace{ \frac{2\neffo\GammaT} {\Gammae}\label{eq:Naddup} }_{\Ncorr}\\ 
    \intertext{and} 
    \Nadddown(\fm) &=
    \underbrace{ \frac{\nth\gammam}{\Gammao}+ \no +  \frac{\ne\Gammae}{\Gammao} }_{\Naddm}
    + \underbrace{ \frac{\neffe\GammaT^2}{\Gammao\Gammae} }_{\Naddtilde}
    - \underbrace{ \frac{2\neffe\GammaT}{\Gammao} }_{\Ncorr}, \label{eq:Nadddown} 
\end{align}
where $\nth$ is the environmental mechanical occupation, $\neffo$ is the effective occupation of the optical cavity due to thermal occupation or the presence of technical noise on the optical pump, and $\no = \neffo+ \nmino$ is the bath occupation that the mechanical mode is coupled to optomechanically, with $\nmino$ denoting the optomechanical backaction limit~\cite{peterson2016laser}.
The quantities associated with the electrical port $\ne$, $\neffe$, and $\nmine$ are defined analogously.
In the resolved-sideband limit, $\no=\neffo$, but we differentiate these noise contributions explicitly here to clarify their physical origins.
In Eqs.~\ref{eq:Naddup} and \ref{eq:Nadddown}, the first three terms comprise noise from motion of the transduction mode $\Naddm$, and the fourth term $\Naddtilde$ is from noise purely of electromagnetic origin measured at the output port.
The fifth term $\Ncorr$ results from destructive interference between the fields responsible for the third and fourth terms.  

\begin{figure}[t!]
    \centering
    \includegraphics{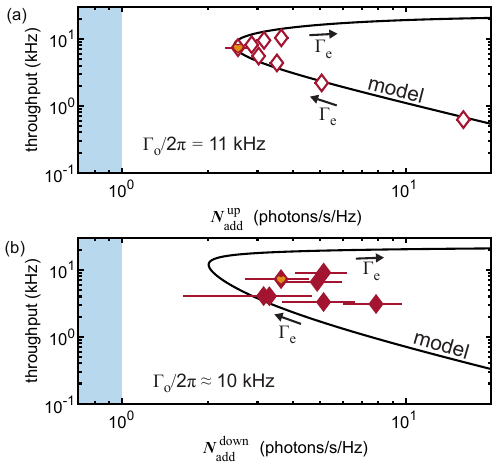}
    \caption{Bidirectional transducer performance. 
    Device performance in upconversion (a) and downconversion (b).
    Maroon diamonds are measurements of our transducer taken while sweeping $\Gammae$ with fixed $\Gammao = 2\pi\cdot11$~kHz in (a), and $\Gammao/2\pi$ roughly fixed in (b), in the range 9-12~kHz.
    Inconsistency in $\Gammao$ between datapoints in (b) is due to drifts in the mode matching between the optical pump and the transducer optical cavity over the approximately two-hour sweep times required for downconversion spectrum measurements.
    Error bars indicate one standard deviation.
    In upconversion, error bars are dominated by statistical uncertainty in the fit parameters used to infer the measurement efficiency and 5\% fractional uncertainty in $\eta$ from separate measurements of $\epsilon_\text{PL}$. In downconversion, error bars are dominated by $\sim 0.2\%$ fluctuations in the noise background of the HEMT amplifier over the 90-second duration of each spectrum measurement, and 10\% fractional uncertainty in $\eta$ from separate measurements of $\epsilon_\text{PL}$.
    Error bars in the $y$-direction are smaller than the data points. 
    The black line is performance modeled at fixed $\Gammao=2\pi \cdot 11$~kHz using independently measured parameters.
    Transducers have nonzero quantum capacity for $\Nadd<1$ (light blue region).
    }
    \label{fig:performance}
\end{figure}

In the limit $\Gammao=\Gammae$, the impedance matching criterion that achieves unit efficiency, Eqs.~\ref{eq:Naddup} and \ref{eq:Nadddown} become the same (for lossless cavities), even in the presence of additional electromagnetic noise at the output port.
However, when $\ne$, $\no$, and $\nth$ differ, often better performance in one direction can be achieved by mismatching the pump strengths. 
Conversely, negligible electromagnetic bath temperatures $\neffo$ and $\neffe$ lead to bidirectional performance for sideband-resolved, high-efficiency transducers.
We note that it is in this noise limit that protocols making use of both ports can mitigate the effects of $\nth$~\cite{higginbotham2018harnessing, rau2022entanglement}.

Practically, the most significant physical effect causing asymmetric transducer performance is that occupations are in general pump-power dependent quantities, and therefore can depend on $\Gammae$ and $\Gammao$:
\begin{equation}
    \nth(\Gammae,\Gammao);\neffe(\Gammae,\Gammao);\neffo(\Gammae,\Gammao).
\end{equation} 
In particular, transduction experiments have observed occupations of the form $\nth(\Gammao)$~\cite{zhao2025quantum,meesala2024non}, $\neffe(\Gammae)$~\cite{brubaker2022optomechanical}, or $\neffe(\Gammao)$~\cite{meesala2024non}. 
Terms of the form $\neffe(\Gammao)$ or $\nth(\Gammao)$ are common because of the high energy of optical photons compared with the lower frequency scales of the microwave and mechanical excitations used in transducers, and 
are ruinous for downconversion performance at high optical power and high duty cycle. 
Though Eqs.~\ref{eq:Naddup} and \ref{eq:Nadddown} apply specifically to doubly-parametric transducers~\cite{brubaker2022optomechanical,zhao2025quantum}, the noise in other architectures also depends on the transducer's thermal environment and coupling rates, and similar expressions can be written. 
The fact that our membrane-based platform does not exhibit noise terms proportional to $\Gammao$ is the reason for its exceptional downconversion noise performance, even though we operate it with high optical circulating powers of approximately 3~mW in this work.

To experimentally probe Eqs.~\ref{eq:Naddup} and \ref{eq:Nadddown} in our device, in Figs.~\ref{fig:performance}(a) and (b) we respectively display upconversion and downconversion results, quantifying $\Nadd(\fm)$ and $\Thr$ while varying $\Gammae$ and with fixed $\Gammao$.
The data points corresponding to the measurements shown in Figs.~\ref{fig:spectrum}, \ref{fig:downconversion}, and \ref{fig:comparison} are marked with hearts. 
The black lines show the predicted performance of our transducer using independently calibrated measurements and modified versions of Eqs.~\ref{eq:Naddup} and \ref{eq:Nadddown} that account for lossy cavities and finite sideband resolution (Appendix~\ref{app: performance model}).
In upconversion, we see agreement between this model and our measurements, up to a systematic difference in the measured value of $\eta$ and that predicted from values of $\Ae$, $\Ao$, $\Gammae$, $\Gammao$, and independently measured cavity loss rates used in the model. 
In downconversion, we see qualitative agreement that reproduces the general trend of our measurement. 

In upconversion, because of its low optical noise and sufficient sideband resolution, our transducer's noise performance is determined by the first two terms in Eq.~\ref{eq:Naddup} (i.e.~$\nth\gammam/\Gammae+\ne$),
as long as $\Gammae$ and $\Gammao$ are reasonably well-matched.
Notably, this approximation implies that $\Naddup$ is independent of $\Gammao$.
As we observe the microwave circuit occupation to follow the form $\neffe = \ae\Gammae + \be$~\cite{brubaker2022optomechanical} (Appendix~\ref{app: technical noise}), the contributions to $\Naddup$ consist of a component of thermomechanical origin that scales as $1/\Gammae$, and a component due to LC circuit occupation that increases linearly with $\Gammae$.
Optimal noise performance in upconversion is achieved by tuning $\Gammae$ to balance these two contributions (Fig.~\ref{fig:performance}(a)).
Relative to our previous work~\cite{brubaker2022optomechanical} (Fig.~\ref{fig:comparison}(a)), the turnaround that results from increasing $\Gammae$ occurs at higher throughput because of a $\gammam$ that is over an order-of-magnitude higher, as measured electromechanically by ringdown of the transduction mode.

Turning to downconversion performance, the first three terms of Eq.~\ref{eq:Nadddown} reveal that, in striking contrast with upconversion, $\Naddm$ can be made quite small simply by increasing $\Gammao$, because the excess motional noise can be routed out the input port by increasing the low-noise optical pump.

If motion of the membrane were the only source of noise, like it is in upconversion, our device would demonstrate quite asymmetrical performance, favoring operation in downconversion with large $\Gammao$.
However, when our device operates in downconversion, the microwave field also contributes to $\Nadd$ through $\Naddtilde$.
Assuming $\GammaT=\Gammao+\Gammae$ to combine the last three terms of Eq.~\ref{eq:Nadddown}, we have
\begin{equation}
    \label{eq:naddDownDetailed}
    \Nadddown(\fm) = \frac{\nth\gammam}{\Gammao}+\no+\frac{\neffe\Gammao}{\Gammae} + \frac{\nmine\Gammae}{\Gammao}. %\nonumber
\end{equation}
Again parameterizing $\neffe= \ae\Gammae+\be$ as above, $\Nadddown$ contains terms proportional to $\Gammao$ and $\Gammao/\Gammae$, so we see that occupation of the microwave circuit prevents arbitrary improvement in downconversion performance with increasing $\Gammao$.
Additionally, the last term in Eq.~\ref{eq:naddDownDetailed} indicates that finite sideband resolution contributing to nonzero $\nmine$ limits the ratio $\Gammae/\Gammao$, causing increased noise at high microwave drive strength.
Optimal downconversion performance is then achieved by tuning $\Gammao$ to balance $\nth\gammam$ against $\ae$, and $\Gammae/\Gammao$ to balance $\nmine$ against $\be$ (Fig.~\ref{fig:performance}(b)).

\section{Conclusions}

We have derived a quantum capacity applicable to transducers across the great variety of physical implementations currently being studied.
The broadband capacity presented here is a practical metric in both the experimental relevance of the model, and in its dependence on the experimentally accessible quantities $\Nadd$ and $\Thr$.
With transduction demonstrations improving in both throughput and added noise, we are approaching broadband capacities at the kebps scale, which is required for transducers to support, on average, the successful transmission of one qubit within every 1~ms lifetime of a superconducting circuit.

Reasonable improvements to our device design would build upon the high upconversion throughput and unprecedented downconversion throughput we show here, bringing kebps-scale broadband quantum capacities within reach.
Refinement of the engineered mechanical isolation of our pinned membrane would suppress the stiff modes in the mechanical spectrum and address the elevated mechanical intrinsic loss rate measured in this device (Appendix~\ref{app: stiff modes}).
Recovering the mechanical coherence we have measured in previous generations of this device, combined with parallel improvement to the electromechanical design to increase the electromechanical coupling rate~\cite{youssefi2023squeezed}, would yield high-throughput quantum-enabled performance in both upconversion and downconversion (Appendix~\ref{app: outlook}).

\section*{Acknowledgments}
We thank Terry Brown, Michael Vissers, and John Teufel for technical assistance, and Georg Arnold, Srujan Meesala, Ravid Shaniv, Benjamin Brubaker, Jonathan Kindem, and Maxwell Olberding for helpful discussions. This work was supported by funding from ARO Grant W911NF2310376, NSF Grant No. PHY-2317149, NSF QLCI Award OMA - 2016244, AFOSR grant FA9550-24-1-0173, NIST, and the Baur-SPIE Endowed Chair at JILA. S.D. additionally thanks support from the DOD through the NDSEG Fellowship Program.

\section*{Data availability}

The experimental data and analysis code used to generate the figures for this work are available at \href{https://doi.org/10.5281/zenodo.15872215}{https://doi.org/10.5281/zenodo.15872215}~\cite{urmey_zenodo_2026}. 
\newline
\newline

\appendix

\section{Integrating the quantum capacity of a thermal loss channel}
\label{app:integrating-capacity}

The right hand side of Eq.~23 of Ref.~\cite{pirandola2017fundamental} gives $\boldsymbol{c}_\text{ub}(\eta, \bar{n})$, which is an upper bound to the two-way quantum capacity $\boldsymbol{c}(\eta, \bar{n})$ of a bosonic thermal-loss channel, modeled as a beamsplitter of transmissivity $\eta$ with a thermal state of $\bar{n}$ mean photons.
While it is not known whether $\boldsymbol{c}(\eta, \bar{n})$ is additive or not, Ref.~\cite{noh2020enhanced} demonstrates that both the coherent information and reverse coherent information of thermal-loss channels is superadditive.
Fortunately, $\boldsymbol{c}_\text{ub}$ is additive, as it is based on the relative entropy of entanglement.
Furthermore, thermal-loss channels are teleportation simulable and hence $\boldsymbol{c}_\text{ub}$ is a strong converse rate~\cite{wilde2017converse}.
Thus any sequence of protocols communicating above this rate must have error probability approaching one, as the number of channel uses increases.

The $\eta$ of Ref.~\cite{pirandola2017fundamental} corresponds to the same $\eta$ used here, and $\bar{n} = \Nadd \frac{\eta}{1-\eta}$, which allows us to reparameterize $\boldsymbol{c}_\text{ub}(\eta, \bar{n})$ as
\begin{widetext}
\begin{equation}
\boldsymbol{c}_\text{ub}(\eta, \Nadd) = 
\frac{1}{1 - \eta}
\Bigl(\eta \Nadd \log_2(\Nadd) -
\big(1 - \eta(1 - \Nadd)\big)\log_2\big(1 - \eta(1 - \Nadd)\big)\Bigr) \, .
\end{equation}
\end{widetext}
We can then integrate $\boldsymbol{c}_\text{ub}$ over a frequency dependent $\eta(f)$ and $\Nadd(f)$.
Since $\boldsymbol{c}_\text{ub}$ is additive, we then get a broadband capacity $\Capac$ with a per-time rate depending on the channel bandwidth.
Assuming that $\eta$ has Lorentzian frequency dependence $\eta(f) = \eta\left[1 + \left(\frac{f}{B/2}\right)^2\right]^{-1}$ and $\Nadd(f) = \Nadd$ is constant across frequencies, we find
\begin{align} \label{eq:integratedCapacityFull}
\Capac(&\Nadd, \Thr) =  
\frac{ \pi\; \Thr/\eta}{\ln(2)}\Bigg[1 - \sqrt{1-\eta(1-\Nadd)} \\
&+\frac{\eta \Nadd}{\sqrt{1-\eta}}
\ln\left(\frac{\sqrt{\Nadd}(1+\sqrt{1-\eta})}
{\sqrt{1-\eta} + \sqrt{1-\eta(1-\Nadd)}}\right)\Bigg]\, . \nonumber
\end{align}
Taking the limit of Eq.~\ref{eq:integratedCapacityFull} as $\eta\to 0$ gives Eq.~\ref{eq:integratedCapacitySmallT}, while taking the limit as $\eta\to 1$ gives $\Capac(\Nadd, \Thr) \approx  \frac{\pi\Thr}{\ln(2)}\left(1 - \sqrt{\Nadd}\right)$. Thus Eq.~\ref{eq:integratedCapacitySmallT} underestimates $\Capac(\Nadd, \Thr)$ by at most a factor of two.

\section{Limiting technical contributions to $\Nadd$}
\label{app: technical noise}

\subsection{Optical pump filtering}
We introduce an additional 130~kHz-linewidth filter cavity to the optical setup detailed in Ref.~\cite{brubaker2022optomechanical}, to filter the output of our Toptica CTL 1050 source laser.
The remainder of the beam preparation remains as previously described~\cite{brubaker2022optomechanical}.
The output of the 130~kHz filter cavity seeds a fiber amplifier, whose output is filtered by a second, 80~kHz-linewidth filter cavity. 
The beam then passes through a double-pass acousto-optic modulator (AOM) before being split to generate the locking, local oscillator, and optical pump beams. 
Using Pound-Drever-Hall locking, the beam is stabilized to the transducer's optical cavity by feeding back to the double-pass AOM and to a piezo on one mirror of the 80~kHz filter cavity.
Two additional PDH feedback loops lock the frequency of the laser to that of the 130~kHz filter cavity, and the frequency of the 130~kHz cavity frequency to that of the 80~kHz filter cavity by actuating a piezo on the 130~kHz filter cavity.
We measure no increase in noise due to the fiber amplifier on the light that is filtered by the 130~kHz cavity.
The additional optical filtering, combined with improved electrical filtering of RF signals used in beam preparation, reduces the phase noise of the optical pump incident on the transducer at a detuning of $\fm$ to -155 dBc/Hz~\cite{urmey2024quantum}, allowing operation of our transducer at $\sim$10~kHz optical dampings with negligible optical contribution to total noise. 

\subsection{Microwave pump filtering}
The effective occupation of the microwave circuit $\neffe$, can have significant contributions from microwave generator phase noise and additional thermal occupation of the circuit, both of which scale with $\Gammae$~\cite{brubaker2022optomechanical}.
In this work, we render the contribution from microwave generator phase noise to be negligible using a room temperature microwave-frequency cavity to filter out the generator phase noise that is on resonance with our LC circuit~\cite{teufel2011sideband}. 
We use a high-$Q$ mode of a cylindrical copper cavity whose frequency is tuned by changing the length of the cavity. The filter cavity provides $>$30~dB of suppression at a detuning of $+\fm$ relative to the carrier. 

\subsection{Residual microwave circuit occupation}
\begin{figure}[t!]
    \centering
    \includegraphics{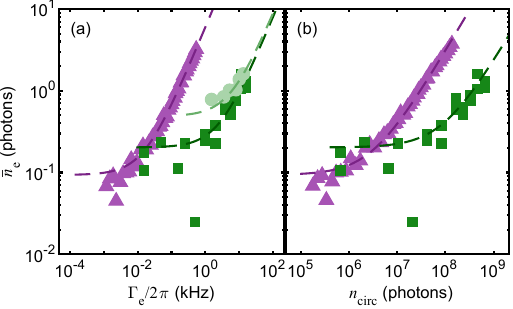}
    \caption{
    Microwave effective mode occupancy. (a) $\neffe$ vs.\ $\Gammae$ as measured electrically (dark green squares) and optomechanically (light green circles), compared with electrical measurement of a prior device (purple triangles)~\cite{brubaker2022optomechanical}. 
    The dashed lines are linear fits to the data. 
    The $\Gammae$-dependent noise is reduced by approximately two orders of magnitude relative to the prior device. 
    (b) We represent the electrically measured data as a function of microwave intracavity photon number $n_\text{circ}$ to account for differences in electromechanical coupling $g_\text{e}$ and microwave loss $\kappa_\text{e}$. 
    The fits shown in (b) are the same as those in (a), appropriately rescaled.
    }
    \label{fig:mwNoise}
\end{figure}

We determine the residual occupation of the microwave mode $\neffe$ by referring the measured microwave noise spectral density $N_\text{det,e}$ back through the microwave measurement chain and dividing by the LC overcoupling ratio.
We compare the measurement of our current device with similar measurements from a previous device~\cite{brubaker2022optomechanical} (dark green squares and purple triangles, respectively, in Fig.~\ref{fig:mwNoise}(a)). 
As a consistency check, we also infer $\neffe$ using sideband asymmetry thermometry of the mechanical mode (light green circles in Fig.~\ref{fig:mwNoise}), optomechanically measured using the optical readout chain, which has higher signal-to-noise ratio than electromechanical measurement in our setup.
We fit these data to the linear expression $\neffe = \ae \Gammae + \be$ (Table~\ref{tab:ne_fit}). 
Relative to our previous device, we observe the power-dependent term $a_\text{e}$ has decreased by approximately two orders of magnitude, enabling a commensurate increase in the transduction bandwidth. 

We attribute the significant performance improvement to two design changes.
Pinning the membrane increased the single-photon electromechanical coupling rate $g_\text{e}$ in this work by approximately 50\%.
We also annealed the \SiN{} film, as described in Ref.~\cite{mittal2024annealing}. 
The annealed \SiN{} film contributed less power-dependent microwave loss and, as we show here, internally generated noise. 
We isolate the internally generated noise in Fig.~\ref{fig:mwNoise}(b), where we plot $\neffe$ as a function of the microwave intracavity photon number $n_\text{circ}$, related to the previous subfigure by $\Gamma_\textrm{e} = 4g_e^2 n_\text{circ}/\kappae$. 
Accordingly, we define the material-dependent noise parameter as $\alpha_\text{e} = 4 g_\text{e}^2 / \kappae \times a_\text{e}$.
The annealed \SiN{} film exhibited a value of $\alpha_\text{e}$ that is smaller than that of the as-deposited film by a factor $13.9$.
We attribute this material improvement to the removal of two-level systems in the \SiN{}.

\begin{table}[h!]
    \caption{Fit to microwave occupancy $\neffe$.}\label{tab:ne_fit}\label{tab:ne_fitV}%
    \begin{tabular}{|@{ }c|c|c|c@{ }|}
        \hline
        device & readout method& $\ae$ $(\text{s})$ & $\be$ \\ 
        \hline
        this work & microwave  & $1.3(1)\times10^{-5}$ &  $0.20(6)$ \\
        this work & optomechanical  & $1.3(4)\times10^{-5}$ & $0.5(3)$ \\ 
        Ref.~\cite{brubaker2022optomechanical}& microwave  & $9.44(5)\times10^{-4}$ & $0.093(6)$ \\
        \hline
    \end{tabular}
\end{table}

\section{Theory of optomechanically stiff modes and the impacts of filtering them out}
\label{app: stiff modes}
Whereas the tensile \SiN{} membrane is light and has low mechanical dissipation, the more massive silicon substrate in which it is embedded supports a far denser distribution of lossier modes.
The mechanical modes of the substrate close in frequency to a membrane mode may hybridize with it, usually resulting in a more membrane-mode-like optomechanically compliant mode, and a more substrate-mode-like optomechanically stiff mode.
We pattern the substrate into a phononic crystal (PNC) with a periodic structure 
to reduce coupling between the membrane and substrate modes and reduce the prominence of stiff modes.
We designed the PNC so that the center of its bandgap would overlap with the transduction mode at 1.5~MHz.

Due to two simultaneous changes to our transducer design, the PNC was less effective at suppressing stiff modes. 
First, annealing the \SiN{} reduced its tensile stress by $\sim$15\%~\cite{jiang2016effect}, which shifted the transduction mode frequency closer to the band edge. 
Second, a structural modification accompanying the membrane pinning design affected the performance of the PnC.
In addition to the pinning post that defines the $\sim$200~nm capacitor spacing between the membrane and the microwave circuit chip, we locate four posts close to the edge of the membrane, along its silicon support frame.
These four edge posts perturb the symmetry of the PNC, reducing its effectiveness at isolating the membrane.
The degraded performance of the PnC combined with the shifted mode frequency led to the the optomechanically stiff modes seen in the noise spectra shown in Figs.~\ref{fig:spectrum} and \ref{fig:downconversion}, and increased $\gammam$ of the transduction mode.
Additionally, one stiff mode appears in $\eta(f)$ in Figs.~\ref{fig:spectrum}(a) and~\ref{fig:downconversion}(a), presenting as a sharp dip in efficiency at frequency $\fs=1.274$~MHz with approximate width of 1~kHz, resulting from the destructive interference of the coherent signal tone propagating through both the stiff mode and the compliant mode.
We corroborate the hypothesized effects of the two mechanical changes using finite-element simulation, and indeed find a higher average density of stiff modes, compared with designs without the additional pinning posts.
Furthermore, we simulate designs that refine the pinned membrane architecture by relocating the edge posts to places of symmetry with respect to the PnC pattern. 
These simulations predict a significant reduction of these detrimental effects.

The relatively sharp frequency dependence of the noise spectrum in this device motivates us to define an integrated noise metric that would be relevant for signals taking advantage of the full transduction bandwidth.
The minimum noise added to a signal with a given frequency profile can be calculated by computing the overlap integral of the added noise spectrum and normalized signal. 
We define the average input-referred added noise in upconversion by integrating over the transduction bandwidth: $\Naddintegrated= \int \Naddup(\omega)\cdot \etaapp(f) df/\left[\int  \etaapp(f) df\right]$.

To recover a weak signal with spectral overlap with the noisy stiff modes, it is possible in principle to notch out the corresponding frequency bands in detection.
This would reduce $\Naddintegrated=7.65$ by approximately a factor of three to $\Naddintegrated=2.64$, by sidestepping hundreds of photons of noise contributed within a narrow bandwidth.
The cost of such filtering is a reduction in fidelity of signal due to the portion that is lost, effectively reducing the efficiency of the transducer by an amount $\eta_\text{filter}$.
Pitching fast single-photon signals at such a filter would also lead to undesirable ringing, a smearing out of the signals in time, which would be an additional source of inefficiency as well as contribute additional noise to subsequent signals. 

To estimate the impact of the drawbacks of implementing a sharp filter at the output of our transducer, we consider a simple protocol of sending single photon signals separated by a repetition time of $t_\text{rep}=3/\GammaT$.
We consider a Lorentzian with linewidth $\GammaT/2\pi=21.7$~kHz, with the regions indicated by the excluded regions in Fig.~\ref{fig:spectrum}~(c) removed from the frequency spectrum to create a ``notched Lorentzian'' frequency profile, and numerically compute its inverse Fourier transform.
Whereas 95\% of the transmitted portion of a fast signal filtered through an ideal Lorentzian would be detected within the first $t_\text{rep}$ of measurement time, 86\% of the same signal would be detected in the case of the notched Lorentzian, due to a combination of signal loss from the notched regions of the filter (94\% efficiency), as well more of the signal extending beyond $t_\text{rep}$ from the ringing of the filter (91\% efficiency). 
The latter effect would also contribute on average an additional 0.09~photons/s/Hz of added noise, assuming single photon signals sent at the full repetition rate of this example of once every $t_\text{rep}=3/\GammaT$.

\section{Transducer performance model}
\label{app: performance model}

To model our transducer's noise performance, we use modified versions of Eqs.~\ref{eq:Naddup} and \ref{eq:Nadddown} that include transducer gain from finite sideband resolution and nonzero internal cavity loss rates $\kappa_\text{e,int}=\kappae-\kappaeext$ and $\kappa_\text{o,int}=\kappao-\kappaoext$.
Additionally, there is a contribution to $\nm$ due to the optomechanical backaction of the locking beam used to PDH lock the transducer optical cavity's resonant frequency $n_\text{lock}\gamma_\text{lock}$. 
From sideband asymmetry thermometry, we find $n_\text{lock}\gamma_\text{lock} +\nth\gammam = 2\pi\cdot2.9$~kHz.
In downconversion, we have
\begin{align}
    \Nadddown(\fm&) = \frac{\nth\gammam+n_\text{lock}\gamma_\text{lock}+\no\Gammao+\ne\Gammae}{\Ao\epsilon\frac{\kappaoext}{\kappao}\Gammao} \nonumber \\
    &+\frac{\neffe\kappaeext\GammaT^2}{\mathcal{A}\etaM\kappae\Gammae\Gammao}
    -\frac{2\neffe\GammaT^2}{\Ao\epsilon\frac{\kappaoext}{\kappao}\Gammao(\Ae \Gammae +\Ao \Gammao)}.
\end{align}
The expression used for $\Naddup$ is analogous, and only includes the first term as the contribution from $\neffo$ is immeasurably small due to the improvements in our filtering of the optical pump. 
The microwave circuit occupation $\neffe(\Gammae)$ is obtained in downconversion and upconversion from the respective fits in Fig.~\ref{fig:mwNoise} to the electrically measured and optomechanically inferred values of $\neffe$.

\section{Operating using a Helium Battery in a cryogen-free dilution refrigerator}
\label{app: helium battery}
We use a BlueFors LD400 dilution refrigerator outfitted with the Helium Battery option to allow us to turn off the pulse tube for approximately 2 hours of continuous low-vibration measurement time.  
During this time, the 50~K temperature stage gradually heats up to approximately 100~K, and the resulting thermal expansion causes the optical mode matching factors to change by approximately 20\%. 
To account for this, we periodically measure the mode matching factors, and interpolate these measured values in analyzing our optomechanical measurements. 

Uncertainty from time-dependence of mode matchings leads to uncertainty in our calibration of $\eta \varpropto \sqrt{\epsilon_\text{PL}}$, and therefore also $\Naddup$ (5\% relative uncertainty) and $\Nadddown$ (10\% relative uncertainty). 
The measured value of $\epsilon_\text{CL}$ is also used in $\Naddup$ through our calibration of $\xi_\text{o}$. 
We use sideband asymmetry thermometry to calibrate each optomechanical spectrum in phonon units, and then divide out the contribution from $\epsilon_\text{CL}$, which is included in our transducer's efficiency. 
Error in the instantaneous value used for $\epsilon_\text{CL}$ contributes to the statistical uncertainty in the time-independent $\xi_\text{o}$, and is reflected in the error bars in Fig.~\ref{fig:performance}(a).

In comparison with the helium precooled dilution refrigerator used in Ref.~\cite{brubaker2022optomechanical}, we observe increased vibration of the optical cavity that we attribute to bubbling of the liquid helium in the battery. 
This vibration drives a $\sim$20~kHz vibrational mode of our optical cavity that the PDH lock to our laser can only partially suppress, resulting in increased optomechanical backaction from the locking beam, due to fast fluctuations of the locking beam detuning.

\section{Microwave readout efficiency calibration}
\label{app: mw readout cal}

Other than the introduction of the tunable filter cavity, the electromechanical readout chain is comparable to that used in Ref.~\cite{brubaker2022optomechanical}, with the superconducting qubit system and the associated circulator removed, and attenuator values on the microwave input reduced to achieve higher microwave pump powers.
Whereas the optomechanical readout efficiency can be accurately calibrated using sideband asymmetry thermometry of the compliant mode, this technique is impractical for calibration of the microwave chain due to the difficulty of measuring the cavity-suppressed lower sideband with its lower signal-to-noise ratio.
Instead, we take advantage of the high signal-to-noise ratio of the electromechanical signal from the transducer's stiff modes, and the simultaneous optomechanical measurement of that motion.
Because the compliant and stiff modes result from hybridization of the same mode with pure membrane motion, the ratio between the optomechanical and electromechanical couplings is identical for these two modes. 
The microwave readout efficiency can thus be expressed as  
\begin{equation}
\xi_\text{e}=\xi_\text{o}\epsilon_\text{CL}\frac{\hstiffe}{\hstiffo}\frac{\kappae}{\kappaeext}\frac{\kappaoext}{\kappao}\frac{\Gammao}{\Gammae} \frac{\Ao}{\Ae}, \nonumber
\end{equation}
where $\hstiffe$ and $\hstiffo$ are the respective electromechanical and optomechanical shot-shot noise normalized spectra at $\fs=1.274$~MHz that are overwhelmingly dominated by motion from the stiff mode at that frequency, and $\epsilon_\text{CL}$ is the factor quantifying the mode matching factor between the transducer's optical cavity and the local oscillator beam used for optical heterodyne detection. 
This method yields $\xi_\text{e} = 0.0086(3)$, approximately consistent with an independent measurement of $\xi_\text{e} = 0.01$ using measurement of the cryogenic insertion loss of the microwave input lines, and measurement of a signal reflecting promptly from the transducer's microwave resonator with detunings of $\pm10$~MHz. 

\section{Device parameters}
We summarize important device parameters for our transducer in Table~\ref{tab:device params}.
Variation in the electromechanical and optical gain factors $\Ae$ and $\Ao$ (and the corresponding backaction limits $\nmine$ and $\nmino$) is from different pump detunings used during individual measurements. 
The range of the mode matching factors includes variation between measurement runs, in addition to the change during a given run described in Appendix~\ref{app: helium battery}.
\begin{table}[h!]
    \centering
    \caption{Device parameters.}\label{tab:device params}
    \begin{tabular}{|@{ }c|c@{ }|}
        \hline
        Parameter &  Value \\
        \hline
        Microwave resonator loss rate, $\kappae$ & $1.06$~MHz\\
        Microwave resonator overcoupling ratio, $\kappaeext/ \kappae$ & $0.67$\\
        Optical cavity loss rate, $\kappao$ & $0.94$~MHz\\
        Optical cavity overcoupling ratio, $\kappaoext/\kappao$ & 0.92\\
        Pump/cavity mode matching factor, $\epsilon_\text{PC}$ & 0.51--0.72\\
        Cavity/LO mode matching factor, $\epsilon_\text{CL}$ & 0.63--0.93\\
        Pump/LO mode matching factor, $\epsilon_\text{PL}$ & 0.36--0.77\\
        Matched transduction efficiency, $\etaM$ & 0.40--0.48\\
        Microwave measurement efficiency, $\xi_\text{e}$ & $0.009$\\
        Optical measurement efficiency, $\xi_\text{o}$ & $0.4$--$0.55$\\
        Transduction mode mechanical frequency, $\fm$ & $1.275$~MHz\\
        Electromechanical gain factor, $\Ae$ & $1.05$--$1.07$\\
        Optomechanical gain factor, $\Ao$ &  $1.03$--$1.05$\\
        Const. mechanical heating rate, $\nth\gammam+n_\text{lock}\gamma_\text{lock}$ & $2\pi \cdot 2.9$ kHz\\
        Electromechanical backaction limit, $\nmine$ & 0.05--0.07\\
        Optomechanical backaction limit, $\nmino$ & 0.03--0.05\\
        \hline
    \end{tabular}
\end{table}

\section{Transduction performance of prior work}
\label{app: prior work}

We summarize the cited work that appears in Fig.~\ref{fig:comparison} in Table~\ref{tab:upconversion points}.
In addition to the works reported in the literature compilation in Ref.~\cite{zhao2025quantum} (Refs.~\cite{mirhosseini2020superconducting, brubaker2022optomechanical, sahu2022quantum, jiang2023optically, meesala2024non, weaver2024integrated, zhao2025quantum}), we include four additional works in upconversion (Refs.~\cite{higginbotham2018harnessing, kumar2023quantum, xie2025scalable, werner2026electro}), and add the measured downconversion performance for Refs.~\cite{sahu2022quantum} and~\cite{jiang2023optically}. 
We include the values of $\eta$, $B$, and $D$ used to compute $\Thr$ for the additional works in Table~\ref{tab:upconversion points}.
Although Refs.~\cite{meesala2024non} and \cite{jiang2023optically} were conducted in the two-mode-squeezing configuration, we include them in Fig.~\ref{fig:comparison}(a) as they provide enough detail to attribute direct transduction performance, making the assumption that noise performance would be equivalent if the blue-detuned pump generating two mode squeezing were instead red-detuned for a beamsplitter interaction (See Supplement of Ref.~\cite{zhao2025quantum}). 
However, the downconversion noise of Refs.~\cite{meesala2024non} and \cite{jiang2023optically} in the two-mode-squeezing configuration is sufficiently high that it falls outside of the plotted range.
The downconversion point included in Fig.~\ref{fig:comparison}(b) for Ref.~\cite{jiang2023optically} was measured in the direct transduction mode.
\begin{table}[h!]
    \centering
    \caption{Transducer performance.}\label{tab:upconversion points}
    \resizebox{\columnwidth}{!}{
    \begin{tabular}{|@{ }c|c|c|c|c|c|c|c@{ }|}
        \hline
        Citation &  $\Nadd$& $\Thr$ (kHz) & $\eta$&  $B$ (kHz)&  $D$& $\epsilon$ \\
        \hline
        Upconversion &&&&&&\\
        \cline{1-1}
        This work & 2.6 & 7.4& 0.34 & 21.7 & 1&0.68\\
        Higginbotham 2018 \cite{higginbotham2018harnessing} &  34&1.6 & 0.47&  3.5&  1&  0.87\\
        Mirhosseini 2020 \cite{mirhosseini2020superconducting} & 0.57 &$4.9\!\times\!10^{-5}$& &&&0.65 \\
        Brubaker 2022 \cite{brubaker2022optomechanical} & 3.2 & 103 & 0.47 & 0.22 & 1 & 0.79\\
        Sahu 2022 \cite{sahu2022quantum} & 0.4 & $3.2\!\times\! 10^{-3}$ & & & & 0.6 \\
        Kumar 2023 \cite{kumar2023quantum}&0.6&0.13 & 0.025&  360&  0.015& -\\
        Jiang 2023 \cite{jiang2023optically} & 1.9 &1.1&&&& -\\
        Meesala 2024 \cite{meesala2024non} &  0.173&$7.5\!\times\! 10^{-4}$ &  &   &   & 0.29\\

        Weaver 2024 \cite{weaver2024integrated}& 6.2 & $2.2\!\times\!10^{-3}$ &&&&0.42\\
        Xie 2025 \cite{xie2025scalable}&  1.24&$2.3\!\times\! 10^{-4}$&  $2.3\!\times\!10^{-3}$&  500&  $2\!\times\!10^{-4}$&0.3\\
        Zhao 2025 \cite{zhao2025quantum} & 0.58 & 0.16 &&&& 0.35\\
        Zhao 2025 \cite{zhao2025quantum} & 9.4 & 6.7 &&&& 0.35\\
        Werner 2026 \cite{werner2026electro}& 0.012 &$5.7\!\times\! 10^{-5}$& $2.2\!\times\! 10^{-4}$& 1300 &$2\!\times\! 10^{-4}$&0.44\\
        Werner 2026 \cite{werner2026electro}& 0.12 &$1.1\!\times\! 10^{-3}$&$2.2\!\times\! 10^{-4}$ & 1300 & $4\!\times\! 10^{-3}$ &0.44\\
        \cline{1-1}
        Downconversion&&&&&&\\
        \cline{1-1}
        This work & 3.6&7.3 & 0.383 & 19.0 & 1&0.76\\
        Sahu 2022 \cite{sahu2022quantum} &  1.1&$7.8\!\times\! 10^{-4}$ & 0.144&  $18000$& $3\!\times\! 10^{-7}$&0.6\\
        Jiang 2023\cite{jiang2023optically} & 99&75 & 0.05 & 1500 & 1 &- \\
        \hline
        \end{tabular}
        }
\end{table}

In an effort to practically and consistently summarize the throughput results of the various experimental efforts, we make the conservative choice to include optical mode matching efficiency losses $\epsilon$, because incorporation into an optical network would require getting signals into and out of the transducer devices.
The values of $\Thr$ listed in Table~\ref{tab:upconversion points} for Refs.~\cite{mirhosseini2020superconducting, brubaker2022optomechanical, sahu2022quantum,jiang2023optically, meesala2024non, weaver2024integrated, zhao2025quantum} are those from Ref.~\cite{zhao2025quantum} multiplied by a factor of $\epsilon$ (see the Supplement to Ref.~\cite{zhao2025quantum} for the calculation of $\Thr$ without the contribution from $\epsilon$ for these works).
Though the contribution from $\epsilon$ is included in $\eta$ and $\Thr$, for clarity, we separately list its value in Table~\ref{tab:upconversion points}.
Note that an optical mode matching factor was not reported in Ref.~\cite{kumar2023quantum}, so that work's throughput would be reduced from additional inefficiency incurred in coupling the device to an optical network. 
In a similar vein, the comparable measurement in Ref.~\cite{jiang2023optically} combines mode matching and loss within the dilution refrigerator.  
For a favorable treatment of that work, we omit their measured value of 0.25, and note here that a similar device from the same group achieved a fiber-to-chip coupling of 0.65~\cite{jiang2020efficient}.
Resonator losses are included in calculating $\eta$ for all works.
\vspace{1 pt}

\section{Outlook toward high-throughput quantum transduction with membrane-based transducers}
\label{app: outlook}

Recovering the environmental heating rate $\nth\gammam = 2\pi\cdot100$~Hz of our prior work \cite{brubaker2022optomechanical} (see Appendix~\ref{app: stiff modes}), combined with improvement of the electrical resonator would allow quantum-enabled, bidirectional transduction at kHz-level throughput. 
By increasing $g_\text{e}$ by a factor of 1.5--2, and assuming a residual microwave circuit occupation $b_\text{e}$ in the range 0.09--0.2, we model that $\Naddup(\fm)$ and $\Nadddown(\fm)$ would each reach 0.6--0.9, at $\Thr=6$~kHz and $\Thr=11$~kHz, respectively.
Given a flat $\Nadd(f)$ spectrum in upconversion, this range corresponds to a value of $\Capac$ of 140~ebps to 1.5~kebps. 
In downconversion, while the value of the nonzero capacity would depend on the form of the noise spectrum, the kHz-level throughput would allow for favorable averaging times in which to characterize such an unprecedented quantum channel.

\bibliography{bib_new}

\end{document}